\documentclass[12pt,preprint]{aastex}

\shorttitle{Volatile Transport inside Super-Earths}
\shortauthors{Levi, Sasselov, and Podolak}

\usepackage{graphicx}
\usepackage{textcomp}
\usepackage{amssymb}
\usepackage{natbib}
\begin{document}
\pagenumbering{arabic}

\title{Volatile Transport inside Super-Earths by Entrapment in the Water Ice Matrix}

\author{A. Levi$^{1,2}$, D. Sasselov$^2$ , and M. Podolak$^1$}\affil{$^1$Dept. of Geophysics \& Planetary Science, Tel Aviv University, Tel Aviv, Israel 69978\\$^2$Harvard-Smithsonian Center for Astrophysics, 60 Garden Street, Cambridge, MA 02138, USA}

\footnotetext{Corresponding author A. Levi, e-mail: amitlevi.planetphys@gmail.com}

\maketitle

\section*{ABSTRACT}

Whether volatiles can be entrapped in a background matrix composing planetary envelopes and be dragged via convection to the surface is a key question in understanding atmospheric fluxes, cycles and composition.
In this paper we consider super-Earths with an extensive water mantle (i.e. water planets), and the possibility of entrapment of methane in their extensive water ice envelopes. 
We adopt the theory developed by \cite{waalplat} for modelling solid solutions, often used for modelling clathrate hydrates, and modify it in order to estimate the thermodynamic stability field of a new phase, called methane filled ice Ih.
We find that in comparison to water ice VII the filled ice Ih structure may be stable not only at the high pressures but also at the high temperatures expected at the core-water mantle transition boundary of water planets.


\section{INTRODUCTION}
The discoveries and characterization of planetary systems orbiting other stars has entered an exciting period when we are starting to gain access to observing the atmospheres of planets that are essentially solid in nature - high-density rocky or icy planets of 1 to 10 Earth masses. These planets, called collectively super-Earths, have been discovered in relatively large quantities, though only a handful have measured radii and masses so far \citep[][and references therein]{Carter2012}. The mean densities derived for these exoplanets reveal a range of possible bulk compositions, ranging from rocky with high iron content \citep[e.g., Kepler-10b,][]{batalha2011} to mini-Neptunes with high H/He fraction \citep[e.g., Kepler-11d,e,][]{lissauer2011}. One of these planets orbiting a nearby M-dwarf star, GJ1214b \citep{charbonneau2009}, has been accessible to spectroscopic studies of its atmosphere with inferences to its composition \citep[e.g.,][and references therein]{bean2011,berta2012}.

At intermediate densities many of the super-Earths should represent a new type of planet, unknown in our Solar System, composed of a rocky core and a water envelope that exceeds 10\% by mass. We will call them water planets; this paper studies the transport of volatiles inside them. Possible examples of water planets could be Kepler-11b, 18b, or 20b.

Massive water planets were introduced conceptually by \cite{kuchner2003} and \cite{leger2004}. \cite{fu10} developed models of the interior dynamics of the water layers of water planets and concluded that materials released from the silicate-iron core should reach the surface, despite the high pressures at the core-mantle boundary and phase transitions. Materials would traverse the water mantle, composed of high pressure ice phases, and reach the surface on timescales of 0.1 to 100 Myr. These are convective overturn timescales: what are the actual materials (e.g., gases) that could be transported to the surface and affect atmospheric observables needs to be determined by a detailed study. Here we attempt a first step in this direction by considering entrapment of volatiles under the extremely high pressures inside water super-Earths. 

The entrapment in water ice of volatile gases, which are mostly hydrophobic species, occurs in structures known as clathrate hydrates: the guest molecules occupy cages of water molecules stabilized predominantly by repulsive interactions between them. Typical clathrate hydrates include gases such as methane, carbon dioxide, oxygen, and nitrogen. Volatile transport by clathrates has been studied in the context of Titan and the icy satellites in our solar system \citep{tobie06,halevi08,schubert10,lunine10,sohl10}.  Because clathrates are of practical terrestrial, as well as of astrophysical importance, much effort has been devoted to measuring their properties in the laboratory \citep[see, e.g.][and the references therein]{fortes10}.  Most experiments are carried out at pressures up to an order of $100$\,MPa, which is high enough to provide useful information for modeling Titan-sized bodies. 

Experiments at higher pressure show that methane clathrates undergo a transition to a more compact form called {\it filled ice} at pressures of around 2\,GPa \citep{loveday01} that can survive at even higher pressures above 86\,GPa \citep{hirai06}. For comparison, in a typical water super-Earth \cite{fu10} estimate that the pressures at the boundary between the ice mantle and the silicate core will be of the order of $100$\,GPa while the temperature will be of the order of $10^3$\,K.  

The basic theory of clathrates that was developed by \cite{waalplat} was later applied by \cite{lunsteve85} to situations of astrophysical interest.  Below we suggest how this theory may be extended to higher pressures and temperatures.  We use this theory to estimate the stability regime of filled ice, and discuss the implications of this for volatile transport in super-Earths.

Clathrates are crystals, whose lattice structure forms cells that act as cages for foreign molecules. The empty cage-like structure is usually thermodynamically unstable, but the captured foreign molecules (i.e. guest molecules) help stabilize the clathrate crystal \citep{waalplat}.  Clathrate hydrate is essentially water ice. In this case a framework of groups of four coordinated water molecules creates a cage-like lattice. The hydrogen-bonded water molecules are slightly distorted, however, from the tetrahedral angle of ordinary ice. There are two basic low pressure ($<1$\,GPa) forms of clathrate hydrates, referred to as structure I (SI) and structure II (SII). Each structure is composed of two types of cages, one small and one large. There is an additional hexagonal structure (SH) whose thermodynamic stability regime is roughly in the $1-2$\,GPa range \citep{hirai01}.

\begin{figure}[ht]
\centering
\includegraphics[scale=0.6]{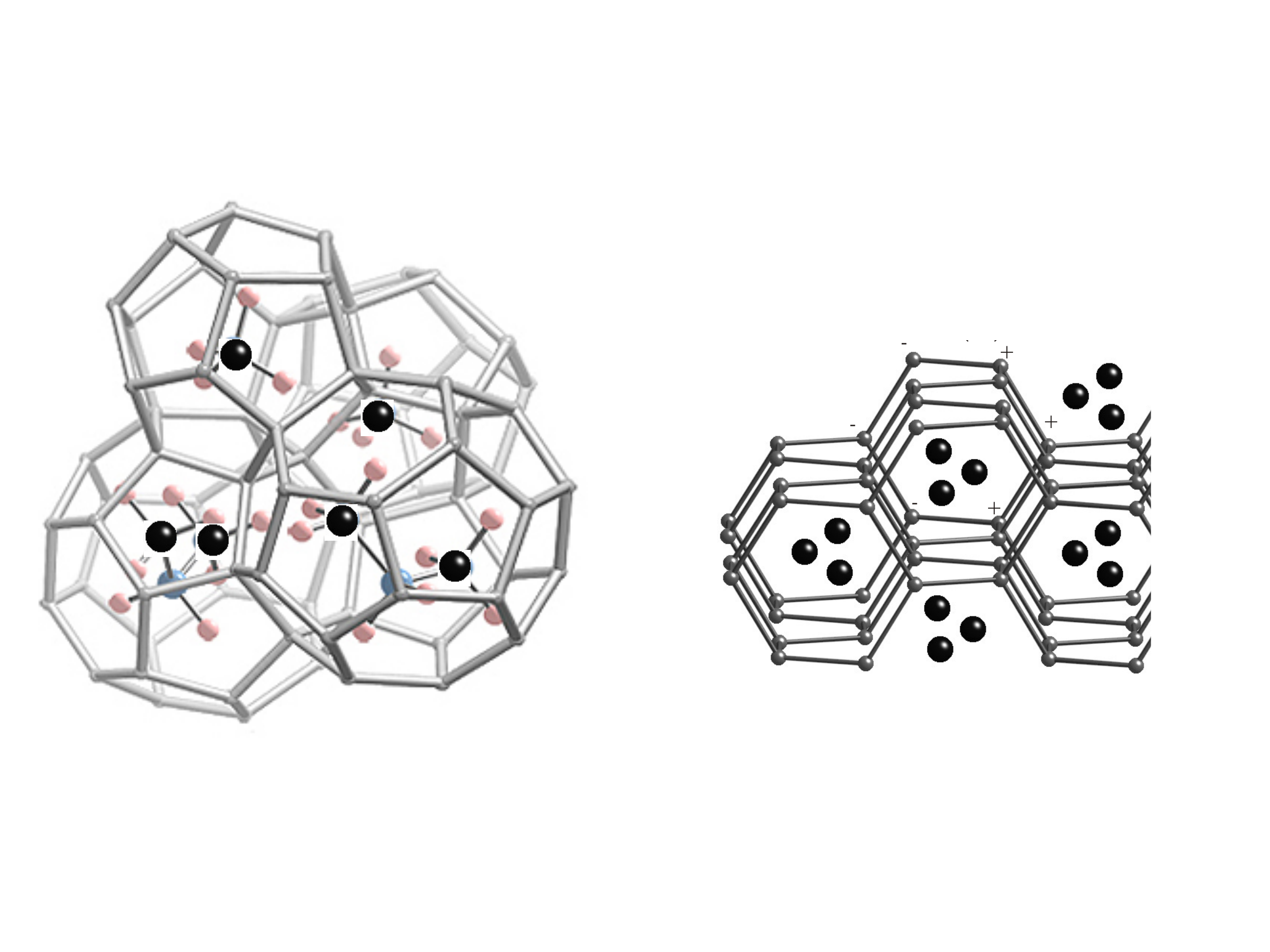}
\caption{\footnotesize{Comparison between methane cage clathrate and methane filled ice Ih. The structure of filled ice Ih is viewed down the c-axis, perpendicular to the widened channels formed in this phase. In water ice Ih every hexagon is hydrogen bonded to its neighbouring hexagons, along the c-axis, in an alternate manner. In methane filled ice Ih three adjacent oxygens of a particular hexagon (plus signs) will bond to one neighbouring hexagon and the other three (minus signs) will bond to the other neighbouring hexagon. This difference results in a widening of the channels perpendicular to the c-axis in filled ice-Ih, therefore allowing for the accomodation of methane. The black balls in the filled ice Ih structure (right panel) denote the entrapped methane molecules. (Filled ice after \cite{loveday01} with permission)}}
\label{fig:comparison}
\end{figure}

Filled water-ice structures also have the capability of trapping volatiles within their hydrogen bonded network, except that instead of the clathrate cages the volatiles are entrapped in channels within the water ice \citep{loveday01}. All cage clathrates have cages whose diameters are much larger than the channels which connect them, while the channels occupied by the guest molecules in filled ice are of constant and smaller diameter along their length (see Figure \ref{fig:comparison}). Their densities also differ significantly: for methane clathrate SI the water-methane ratio is 5.75:1, while for methane filled ice it is 2:1. The increased guest-host and guest-guest intermolecular interaction may help explain the increased stability of the filled ice against high pressure compression and decomposition \citep{machida07}. Diamond anvil experiments, using methane as the guest molecule, suggest such structures may survive pressures beyond $86$\,GPa \citep{hirai06}. 

The ability of clathrate structures to incorporate relatively large amounts of volatile species under conditions where the pure volatiles are highly unstable in the condensed phase is a primary characteristic that makes them so important to solar system and exoplanet studies.
Enclathrating volatiles in icy planetesimals results in their inclusion in planetary icy mantles. As the icy mantle grows its inner layers become pressurized, and at pressures higher than about 2\,GPa even SH clathrate hydrates are destabilized. Experiments show that the rate of pressure increase controls the resulting water phase: if the pressure is raised in steps of $0.2-0.5$\,GPa every $1-2$\,hr, the methane clathrate will decompose to ice VII and pure methane ice.  However, if the sample of methane clathrate hydrate is kept at more than 3\,GPa for $12$\,hr the molecules have enough time to rearrange themselves into a filled ice structure that can sustain very high pressures without decomposing \citep{lovedaynat01}. Thus it appears that the filled ice phase will be an important component of a planetary ice mantle, especially inside water super-Earths.

In this paper we begin by revisiting the thermodynamic stability theory of clathrates and extend it to higher pressures and temperatures. We consider only methane as the guest molecule here. In section 3 we consider the case of methane filled ice. We conclude with numerical results and brief application to super-Earths.

\section{THERMODYNAMIC STABILITY FIELD}

The basic theory of clathrate hydrates views their stability in terms of two phases:  The ordinary water phase (e.g. ice Ih, liquid water, etc. - the $\alpha$ phase) and the empty clathrate hydrate structure (the $\beta$ phase) + a gas of guest molecules. The chemical potentials of the two phases are equal on the thermodynamic equilibrium boundary between them.  For clathrates this may be written as \citep{waalplat}:
\begin{equation}\label{muQ}
\mu_Q^\alpha = \mu_Q^\beta+kT\sum_i\nu_i\ln\left(1-\sum_Ky_{Ki}\right)
\end{equation}
\begin{equation}\label{yKi}
y_{Ki}=\frac{f_KC_{Ki}}{1+\sum_Jf_JC_{Ji}}
\end{equation}
Here the subindex $Q$ refers to the molecule that makes up the host molecular network, which in our case is H$_2$O. The second term on the RHS of Eq.(\ref{muQ}) represents the contribution of the guest molecules to the clathrate hydrate chemical potential. $T$ is the temperature, $k$ is Boltzmann's constant, $\nu_i$ is the ratio between the number of $i$ type cages to water molecules per cubic unit crystal, and finally, $y_{Ki}$ is the probability that a guest molecule of type $K$ occupies a clathrate cage of type $i$. This last function is given in Eq.(\ref{yKi}) in terms of the volatile fugacity ($f_K$) and its Langmuir constant ($C_{Ki}$). We will give further information on the Langmuir constant below and refer the reader to \cite{waalplat} for a more detailed derivation.

Since most high pressure experimental information is for methane-filled water-ices we restrict ourselves to the case of a single species of guest molecule model, and omit the summation over $K$. The equilibrium equations may therefore be written as:
\begin{equation}
\frac{\mu_{H_2O}^\alpha}{kT} = \frac{\mu_{H_2O}^\beta}{kT}+\sum_i\nu_i\ln\left(1-y_{CH_4,i}\right)
\end{equation}
\begin{equation}
y_{CH_4,i}=\frac{f_{CH_4}C_{CH_4,i}}{1+f_{CH_4}C_{CH_4,i}}
\end{equation} 

Recalling that both the fugacity and the Langmuir constant are functions of temperature and pressure, we differentiate and combine the last set of equations to get:
$$
\frac{\partial}{\partial T}\left(\frac{\mu_{H_2O}^\alpha}{kT}-\frac{\mu_{H_2O}^\beta}{kT}\right)_PdT+\frac{1}{kT}\frac{\partial}{\partial P}\left(\mu_{H_2O}^\alpha-\mu_{H_2O}^\beta\right)_TdP=
$$
\begin{equation}\label{dPdT}
-\sum_i\nu_i\frac{1}{1+f_{CH_4}C_{CH_4,i}}\left[\frac{\partial}{\partial P}\left(f_{CH_4}C_{CH_4,i}\right)_TdP+\frac{\partial}{\partial T}\left(f_{CH_4}C_{CH_4,i}\right)_PdT\right]
\end{equation}

We introduce the thermodynamic relations:
\begin{eqnarray}
\frac{\partial}{\partial T}\left(\frac{\mu_{H_2O}}{kT}\right)_P\equiv -\frac{H_{H_2O}}{kT^2}\\
\frac{1}{kT}\frac{\partial}{\partial P}\left(\mu_{H_2O}\right)_T\equiv \frac{V_{H_2O}}{kT}
\end{eqnarray}
Where $H_{H_2O}$ and $V_{H_2O}$ represent the enthalpy and volume per water molecule, respectively. Inserting these thermodynamic relations into Eq.(\ref{dPdT}) and solving for $dP/dT$ gives:
\begin{equation}\label{dpdt2}
\frac{dP}{dT}=\frac{\frac{1}{kT^2}\left(H_{H_2O}^\alpha-H_{H_2O}^\beta\right)-\sum_i\nu_i\frac{1}{1+fC_i}\left(\frac{\partial f}{\partial T}C_i+f\frac{\partial C_i}{\partial T}\right)}{\frac{1}{kT}\left(V_{H_2O}^\alpha-V_{H_2O}^\beta\right)+\sum_i\nu_i\frac{1}{1+fC_i}\left(\frac{\partial f}{\partial P}C_i+f\frac{\partial C_i}{\partial P}\right)}
\end{equation}

Here we have omitted the index CH$_4$ for convenience. For a homogeneous substance (i.e. composed only of water molecules) the chemical potential equals the Gibbs free energy per particle, $\tilde{G}$, which, combined with the assumption of constant temperature gives:
\begin{equation}
d\mu=d\tilde{G}=VdP=kTd\ln P
\end{equation}
Here all extensive parameters are per particle. For the final equality on the RHS we have used the equation of state for an ideal gas, so that the last relation between pressure and Gibbs free energy is only valid for this case. When the gas is not ideal, we can retain the functional form if we replace the pressure with the fugacity, $f$, which acts as an effective pressure function to correct for the effect of intermolecular interactions and which obeys the following relation \citep{smithness}:
\begin{equation}
d\mu=d\tilde{G}=VdP\equiv kTd\ln f
\end{equation}
From the last relation we have:
\begin{equation}\label{dfdP}
\left(\frac{\partial f}{\partial P}\right)_T=\frac{Vf}{kT}
\end{equation}
Where $V$ is the volume per methane molecule. For high pressures we can solve Eqs.(\ref{muQ}) and (\ref{yKi}) for the case of an SI methane clathrate hydrate numerically. We find that $fC_i\sim 10^3\gg 1$.  A similar result was found by \cite{lunsteve85}. Inserting this numerical result together with Eq.(\ref{dfdP}) into Eq.(\ref{dpdt2}) yields:
\begin{equation}\label{13}
\frac{dP}{dT}=\frac{\frac{1}{kT^2}\left(H_{H_2O}^\alpha-H_{H_2O}^\beta\right)-\sum_i\nu_i\left(\frac{\partial}{\partial T}\ln f+\frac{\partial}{\partial T}\ln C_i\right)}{\frac{1}{kT}\left(V_{H_2O}^\alpha-V_{H_2O}^\beta\right)+\sum_i\nu_i\left(\frac{V}{kT} +\frac{\partial}{\partial P}\ln C_i\right)}
\end{equation}

The definition of the Langmuir constant according to \cite{waalplat} is:
\begin{equation}
C_{Ki}\equiv\frac{h_{Ki}}{kT\zeta_{K}(T)}
\end{equation}
where $h_{Ki}$ is the single cell canonical partition function for a $K$ type guest molecule in an $i$ type cage. $\zeta_K$ is the quantum number density function for a $K$ type molecule in an ideal gas and is independent of pressure. The cell partition function depends on pressure both via the cell dimension and through the form of the guest-host potential, so: 
\begin{equation}
\frac{\partial}{\partial P}\left(\ln C_i\right)_T=\frac{\partial}{\partial P}\left(\ln h_i\right)_T
\end{equation}  
Here again we have omitted the index $K$. We are left with estimating the derivative of the single cell partition function, for which we give the following quasi-classical form:
\begin{equation}
\frac{\partial}{\partial P}\left(\ln h_i\right)_T=\frac{\partial}{\partial P}\ln\left\lbrace \frac{1}{\hbar^3}\int d^3\textbf{\textit{r}}\int d^3\textbf{\textit{p}}e^{-1/kT[\epsilon_{rot,i}+\epsilon_{vib,i}+\epsilon_{trans,i}+W_i]} \right\rbrace
\end{equation} 
Here we have divided the Hamiltonian of the guest molecule into its separate kinetic and potential contributions. As a first order approximation we assume that the kinetic degrees of freedom of the entrapped molecule are unaffected by its inclusion in the water network. This is a common approximation in clathrate modeling, and experiment shows that in solidified form methane molecules rotate freely as in an ideal gas \citep{hazen80}. In this approximation the kinetic degrees of freedom will contribute a function that is only a function of temperature, and the logarithm will cancel upon differentiation with respect to pressure, thus yielding a simplified form:
\begin{equation}
\frac{\partial}{\partial P}\left(\ln h_i\right)_T\approx\frac{\partial}{\partial P}\ln\int e^{-W_i/kT}d^3\textbf{\textit{r}}
\end{equation}

In clathrate hydrates the cages entrapping the volatiles are assumed spherical \citep[e.g.][]{sloan,mckoysinan} so that the integration is over a spherical cage. In filled ices the cages are actually cylindrical channels within the water ice structure \citep{loveday01} so that we may write:
\begin{equation}
\frac{\partial}{\partial P}\ln\int e^{-W_i/kT}d^3\vec{r}=\frac{\partial}{\partial P}\ln\int_{z1}^{z2}dz\int_0^{2\pi}d\phi\int_0^a e^{-W_i(r,\phi,z)/kT}rdr
\end{equation} 
Where we assume the cylindrical water-made channel has radius $a$ and since intermolecular potentials fall rapidly with increasing distance we limit the integration along the $z$ coordinate to the finite values $z_1$ and $z_2$.

Both the limits of integration and the potential interaction of the methane molecule with its surroundings, $W_i$, may depend on the pressure. 
Raman spectra of methane filled water ice shows increases in the attractive potential between methane and its water network host with pressure increases \citep{machida07}. This was also suggested by \cite{hirai06}. The rearrangement of the water network, from cage clathrate to the filled ice structure, reduces molecular distances by $\sim 0.5\times 10^{-8}$~cm. It is also suggested, from intermolecular distance considerations, that the guest-host Lennard-Jones potential interaction estimated for clathrates is enhanced by weak hydrogen bonds between guest and host in the filled ice structure \citep{loveday01}. Since the guest-host potential energy changes considerably with pressure, a good first approximation will be to consider only the change of $W_i$ with pressure, therefore allowing us to insert the derivative with respect to pressure into the integrand, giving: 
\begin{equation}
\frac{\partial}{\partial P}\left(\ln C_i\right)_T=-\frac{1}{kT}\frac{\int_{z1}^{z2}\int_0^{2\pi}\int_0^are^{-W_i/kT}\left(\frac{\partial W_i}{\partial P}\right)_Td\phi dzdr}{\int_{z1}^{z2}\int_0^{2\pi}\int_0^are^{-W_i/kT}d\phi dzdr}
\end{equation}
 
Finally, taking a spatial average for the partial derivative appearing in the numerator, we have:
\begin{equation}
\frac{\partial}{\partial P}\left(\ln C_i\right)_T=-\frac{1}{kT}\left\langle\left(\frac{\partial W_i}{\partial P}\right)_T\right\rangle
\end{equation}  
Inserting this last relation into Eq.(\ref{13}) yields:
\begin{equation}
\frac{dP}{dT}=\frac{\frac{1}{T}\left(H_{H_2O}^\beta-H_{H_2O}^\alpha\right)+kT\sum_i\nu_i\left(\frac{\partial}{\partial T}\ln f+\frac{\partial}{\partial T}\ln C_i\right)}{V_{H_2O}^\beta-V_{H_2O}^\alpha+\sum_i\nu_i\left\langle\left(\frac{\partial W_i}{\partial P}\right)_T\right\rangle-\sum_i\nu_iV}
\end{equation}
Since the pressure varies by many orders of magnitude and the temperature does not, and since pressure has a dominant effect in determining the correction for the system non-ideality, we assume that: 
$$\frac{\partial C_i}{\partial T}\gg \frac{\partial f}{\partial T}$$
We may further approximate the following:
$$
kT\sum_i\nu_i\frac{\partial}{\partial T}\ln C_i=kT\sum_i\nu_i\frac{\partial}{\partial T}\ln\left(\frac{1}{kT}\int d^3\vec{r}e^{-W_i/kT}\right)
$$
\begin{equation}
=kT\sum_i\nu_i\left[\frac{\int d^3\vec{r}\frac{W_i}{kT^2}e^{-W_i/kT}}{\int d^3\vec{r}e^{-W_i/kT}}-\frac{1}{T}\right]=\sum_i\nu_i\left[\frac{1}{T}\left\langle W_i\right\rangle-k\right]
\end{equation}
Where in the last step we have again averaged the intermolecular interaction of the methane molecule spatially over its surroundings. Inserting the last formula into the relation for $dP/dT$ gives a Clausius-Clapeyron type equation of the form:
\begin{equation}\label{clausclap}
\frac{dP}{dT}=\frac{\frac{1}{T}\left(H_{H_2O}^\beta-H_{H_2O}^\alpha \right)+\sum_i\nu_i\left[\frac{1}{T}\left\langle W_i\right\rangle-k\right]}{V_{H_2O}^\beta-V_{H_2O}^\alpha+\sum_i\nu_i\left\langle\left(\frac{\partial W_i}{\partial P}\right)_T\right\rangle-\sum_i\nu_iV}
\end{equation}   
This modified Clausius-Clapeyron type equation is a generalization of the equation given in \cite{lunsteve85} in order to explain the behaviour of clathrates. We will use this formalism to obtain the thermodynamic field of stability for the filled water ice as well. For this purpose we need to evaluate the different terms appearing in this equation, which we do in the next section.

\section{APPLICATION TO HIGH PRESSURE}

\subsection{Clathrate Hydrate}

Starting with clathrate hydrates, let us examine the numerator of Eq.(\ref{clausclap}). At low pressures and temperatures the $\alpha$ phase is water-ice Ih whose enthalpy is taken to be equal to the enthalpy of the empty hydrate ($\beta$) phase \citep{lunsteve85}. The potential of interaction $W_i$ is attractive (negative) and of a Lennard-Jones type, so generally the numerator is negative.  In the denominator, at low pressures, the gaseous volatile volume ($V$) is the dominant factor appearing with a minus sign. Therefore the derivative $dP/dT$ is positive. The potential of interaction hardly changes with increasing pressure in this low pressure regime.
 
As we increase the temperature, the pressure increases till we reach the melting point for water-ice Ih and the $\alpha$ phase now represents liquid water. Due to the enthalpy of fusion, the enthalpy difference appearing in the numerator is no longer negligible and causes a sharp increase in the absolute value of the numerator. This is manifested as a sharp increase in the derivative $dP/dT$. Every increase in temperature is now accompanied by a steeper increase in pressure. At room temperature at 5\,MPa, methane gas already deviates from ideality enough so that we need to consider a second virial correction; at 10\,MPa a third virial correction is required, and so on \citep{hirschcurbird}. This means that the volume a methane molecule occupies in the gas decreases with pressure.  At high enough pressure the volatile gas contracts enough so that the empty clathrate volume equals the liquid water volume plus the compressed methane volume and the derivative $dP/dT$ diverges.

\begin{figure}[ht]
\centering
\includegraphics[scale=0.6]{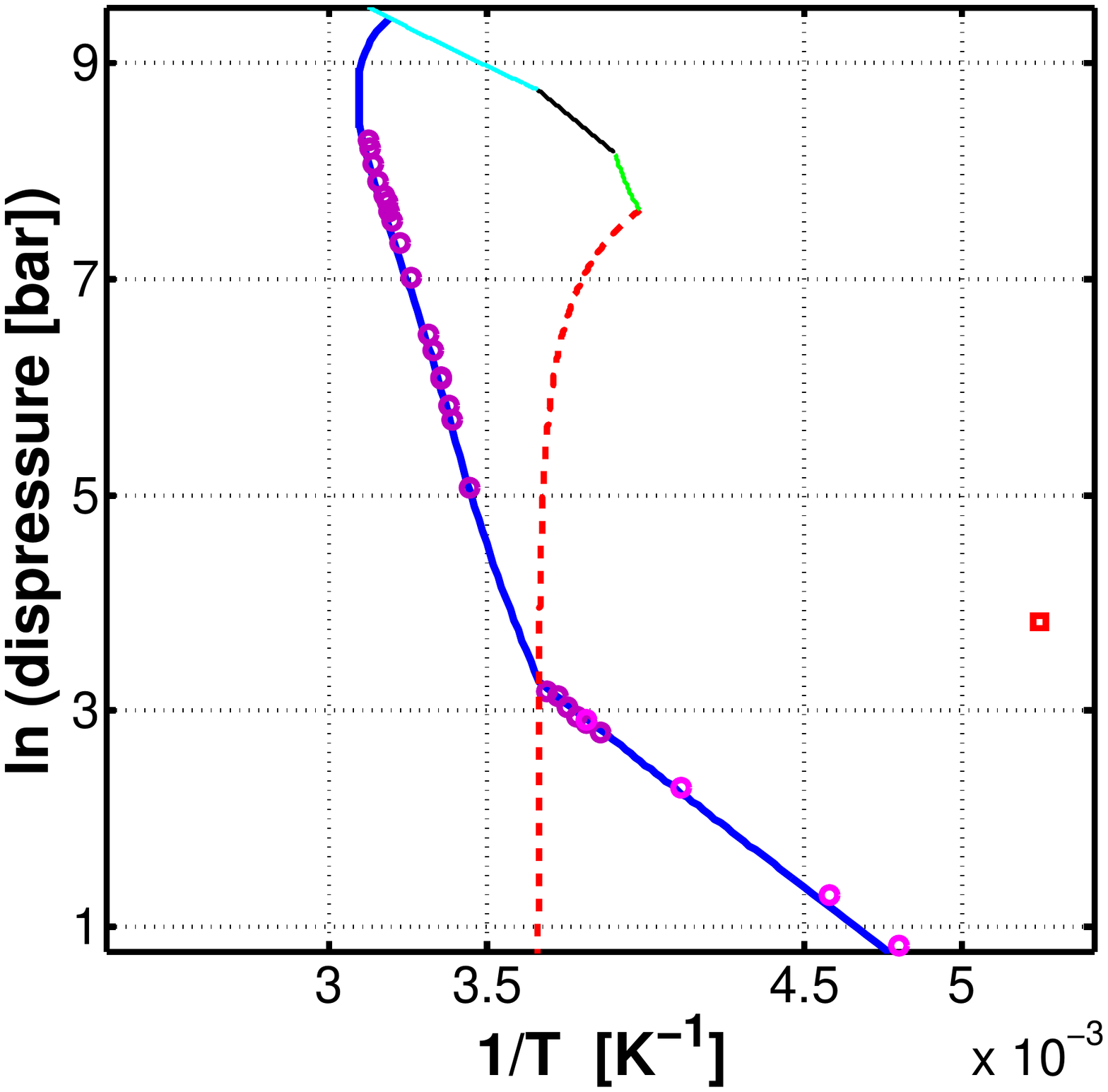}
\caption{\footnotesize{This figure describes the thermodynamic stability field for CH$_4$ clathrate hydrate structure I. The dotted line is water ice Ih melting curve, the thick curve profile (coloured blue in the online version) is the dissociation pressure curve for CH$_4$ clathrate hydrate structure I, which clearly follows the available data points (circles). The other melting curves are for water ice III, water ice V and water ice VI. The square (coloured red in the online version) marks the critical point for Methane.}}
\label{fig:DisPressCH4SI}
\end{figure}

Any further increase in pressure brings about a situation where the volume occupied by a clathrate water molecule is larger than the sum of the volumes occupied by a water molecule in the liquid phase and a methane molecule in the gas phase weighted by the hydration number ($\nu$). The result is that the derivative $dP/dT$ becomes negative and the high pressure stability limit of the clathrate hydrate is attained and the clathrate dissociates.  This general type of clathrate behavior is shown in fig.~\ref{fig:DisPressCH4SI}, which we have derived by numerically solving the set of Eqs.(\ref{muQ}) and (\ref{yKi}).

Now let us consider what happens for the case of filled water-ice. This ice is formed in the laboratory at room temperature and at a pressure of $\sim 2$~Gpa. The structure is an ice Ih water network which is distorted in such a way that the interconnecting channels widen to accommodate the methane molecules within the hydrogen bonded network. This distortion from the usual structure of water ice Ih is manifested by a change in the tetrahedral angles \citep{loveday01}. The $\beta$ phase will now describe the filled ice structure while the $\alpha$ phase represent either water ice VII or fluid water, appropriately. 

To obtain the thermodynamic stability field for filled water ice we shall require the temperature and pressure dependencies for both the different constituents' volumes and for the attractive potential between a methane molecule and its surroundings. In addition, the enthalpy difference between filled ice and fluid water ought be estimated for the case of stability with respect to fluid water.   

\subsection{High Pressure Equations of State}

We start with the general relation:
\begin{equation}\label{generaleos}
\frac{dV}{V}=\chi dT-\kappa dP
\end{equation}
Where $\chi$ and $\kappa$ are the volumetric thermal expansivity and isothermal compressibility respectively. We assume the bulk modulus, $B$, is linearly dependent on pressure and may be written as:
\begin{equation}\label{bulkmodulus}
B\equiv\frac{1}{\kappa}=B_0+\tilde{B_0}P
\end{equation} 
Combining Eqs.(\ref{bulkmodulus}) and (\ref{generaleos}) yields after integration:
\begin{equation}\label{eos1}
V(T,P)=V(T_0,P_0)\left(\frac{B_0+\tilde{B_0}P}{B_0+\tilde{B_0}P_0}\right)^{-1/\tilde{B_0}}exp \left(\int_{T_0}^T\chi(T,P) dT\right)
\end{equation} 
If we keep the temperature constant at $T_0$ we end up with the Birch-Murnaghan equation of state. By setting the reference temperature ($T_0$) to room temperature, we can assign to each solid constituent a proper value for its bulk modulus ($B_0$) and its pressure gradient ($\tilde{B_0}$) by using published room temperature hydrostatic compression experiments. We use the data published in \cite{hemley87} for water ice up to $128$~GPa, for assigning a bulk modulus for ice VII.
For assigning a bulk modulus for solid methane we use the data published in \cite{hazen80} for phase I and in \cite{umemoto2002} for phases A and B up to $37$~GPa. Solid methane transforms at high pressures to a hexagonal close packed structure \citep{bini1997}, this phase was difficult to account for as we found no volumetric data for it, rather we extrapolated from the phase B data in \cite{umemoto2002} to higher pressures. 
For the filled ice structure we used room temperature, experimentally deduced, unit cell volumes up to $42$~GPa by \cite{hirai2003}. 

For the thermal expansivity we have adopted the approach of \cite{fei1993}, who determined the thermal expansivity of water ice VII by fitting their volumetric experimental data to an equation of state of the form given above (Eq.\ref{eos1}). These authors assumed the following dependence for $\chi$ on the pressure:
\begin{equation}\label{expansivity}
\chi(T,P)=\chi(T,P_0)\left(\frac{B_0+\tilde{B_0}P}{B_0+\tilde{B_0}P_0}\right)^{-\eta}
\end{equation} 
Where $\chi(T,P_0)$ is taken to be a linear function of the temperature and $\eta$ is well fitted with a numerical value of $0.9$.
For the thermal expansivity of solid methane we adopt the formalism of Eq.(\ref{expansivity}) for its dependency on pressure and set the expansivity value at $P_0$ to be $10^{-3}$~K$^{-1}$, according to experimental data given by \cite{heberlein1970}.
The thermal expansivity of filled ice is not known. As filled ice is a combination of a hydrogen bonded-network and methane molecules between which there is van der Waals type attraction, we suggest its thermal expansivity to be intermediate between that for ice VII and for solid methane.

For the equation of state of fluid water at high pressure (up to $100$~GPa) we adopt the formalism derived using molecular dynamic simulations by \cite{Belonoshko1991}. Although this formalism is inherently dependent on the type of model used for the water molecules' intermolecular potential it does show good agreement with recent experimental data for high pressure water fluid density \citep{Goncharov2009}.

We shall now turn to evaluate the energy of interaction of a methane molecule with its surroundings in the water filled ice structure.

\subsection{The Interaction Potential $W_i$} 

In order to build the thermodynamic stability field for methane filled water ice, we need to approximate how the potential of interaction, of the methane molecule with its surroundings, depends on the temperature and the pressure. 
As was shown by \cite{raghavendra2008} a hydrogen bond may form between the water electron poor hydrogen and the center of the methane tetrahedral face which is electron rich. It was further shown by these authors that the bond energy ($E_{bond}$) for such an interaction is $-6.3$~kJ~mol$^{-1}$. For comparison a methane-methane van der Waals potential well is about  $-1.2$~kJ~mol$^{-1}$ \citep{hirschcurbird}. It was also suggested by \cite{loveday01} that weak hydrogen bonds are formed in the filled water-ice structure, between the water network and the dissolved methane molecules. 

Let us consider a simple approximate model where with increasing pressure more of the tetrahedral faces, per methane molecule, create such hydrogen bonds with the water network. An increase in temperature will have the opposite effect. That the number of hydrogen bonds per molecule increases with pressure and decreases with temperature is known for water structures \citep[e.g.][]{kalinichev94,pattanayak2011}.
We may therefore write for the spatially averaged potential of interaction of a methane molecule with its surroundings:
\begin{equation}\label{avepotential1}
\left\langle W_i\right\rangle = n(T,P)E_{bond}
\end{equation}
Where $n(T,P)$ is the number of hydrogen bonds between a methane molecule and the water network, at a given pressure and temperature, and is bounded between $0$ and $4$, where in the upper limit all of the methane four tetrahedral faces are hydrogen bonded to the water network. 

We normalize $n$ to give it a probability interpretation (i.e. What is the probability a bond will form at a given pressure-temperature point). We further assume a division into a temperature dependent and pressure dependent probabilities, thus:
\begin{equation}
n(T,P)\equiv n_1(T)n_2(P)
\end{equation}
For the temperature dependent probability we assume a Boltzmann type form, of:
\begin{equation}\label{probT}
n_1(T)=\frac{1}{1+exp(-4|E_{bond}|/kT)}
\end{equation}
To obtain the form for the pressure dependent probability ($n_2$) we use the fact that a formation of a hydrogen bond is accompanied by a substantial penetration of the hydrogen bonding molecules into each other. For the case of the hydrogen bonding between water and methane the combined van der Waals radius of $2.90\times 10^{-8}$~cm, before the bonding, reduces to $2.47\times 10^{-8}$~cm, after bonding \citep{raghavendra2008}. This interpenetration may therefore account for a $40\%$ reduction in the crystal volume. This is considerable enough so that we may relate this interpenetration (to first approximation) to the solid compressibility, $\kappa$. 
At low pressures only a few hydrogen bonds are formed and the solid is easily compressed, as many bonds are still ready to be formed. As the pressure increases more bonds form, per molecule, and it becomes more difficult to further compress the solid. Hypothetically, at a high enough pressure, all bonds per molecule are already formed and it is no longer possible to further compress the solid via the route of hydrogen bonding molecular interpenetration. Adopting this model, to first approximation, we can write for the molecular volume, $V$, the following:
\begin{equation}
V=n_{HB}V_{HB}+n_{nHB}V_{nHB}
\end{equation}
Where $n_{HB}$ and $n_{nHB}$ are the probabilities a methane molecule is fully hydrogen bonded to the water network or not hydrogen bonded at all respectively, and $V_{HB}$ and $V_{nHB}$ are the molecular volumes associated with these two molecular situations respectively. If the number of hydrogen bonds indeed determines, to first approximation, the crystal volume, we may say:
\begin{equation}\label{kappamodel}
\kappa\equiv -\frac{1}{V}\left(\frac{\partial V}{\partial P}\right)_T=\frac{-1}{n_{HB}V_{HB}+n_{nHB}V_{nHB}}\left(V_{HB}\frac{dn_{HB}}{dP}+V_{nHB}\frac{dn_{nHB}}{dP}\right)
\end{equation}  
Since $n_{HB}+n_{nHB}=1$, an integration of Eq.(\ref{kappamodel}) yields:
\begin{equation}\label{probP}
n_2(P)=n_{HB}(P)=\frac{1-exp\left(-\int_0^P\kappa dP\right)}{1-\frac{V_{HB}}{V_{nHB}}}
\end{equation}  
Inserting Eqs.(\ref{probP}) and (\ref{probT}) into Eq.(\ref{avepotential1}) gives for the spatially averaged potential energy, of a methane molecule with its surroundings, the following form:
\begin{equation}
\left\langle W_i\right\rangle \approx \frac{4}{1+e^{-4|E_{bond}|/kT}}\left(1-e^{-\kappa_0 P}\right)E_{bond}
\end{equation}
Its derivative with respect to pressure will therefore be:
\begin{equation}
\left\langle\left(\frac{\partial W_i}{\partial P}\right)_T\right\rangle \approx \frac{4\kappa_0 E_{bond}}{1+e^{-4|E_{bond}|/kT}}e^{-\kappa_0 P}
\end{equation}   
Now that we have the approximate temperature and pressure dependencies for the terms appearing in Eq.(\ref{clausclap}) we may integrate it numerically to obtain the stability regime for the filled ice structure.

\subsection{Numerical Results and Discussion}

The numerical integration of Eq.(\ref{clausclap}) is shown in fig.~\ref{fig:phasediagram}. As we have mentioned above, we consider it reasonable to assume the thermal expansivity of the filled ice to be intermediate, between that for solid methane (a van der Waals solid) and water ice VII. From the laboratory data (see subsection $3.2$) we know the thermal expansivity for solid methane is an order of magnitude larger than that for ice VII. We integrate Eq.(\ref{clausclap}) assuming thermal expansivity for the filled ice two times, three times and five times larger than the experimental value for ice VII. As a limiting case we also solve assuming filled ice has a thermal expansivity equal to that of water ice VII.
The four curves are given in the figure.
It is appropriate to note here that \cite{sloan} gives for a SI clathrate hydrate (also a combination of methane and water) a thermal expansivity some five times larger than the thermal expansivity for ice Ih. More recent experiments confine the latter ratio to be between two to four \citep{hester2007}. 

In addition, the parameter $\nu$, the hydration number, is important to Eq.(\ref{clausclap}), as defined earlier. Generally speaking, this parameter represents the ability to include volatiles in the water network. As the ratio between water and methane in the filled ice is $2:1$ \citep{loveday01} then filled ice modeling requires a value of $1/2$ for $\nu$. 
Given that, we find the packing efficiency for filled ice to be greater than for the case of separation to pure solid constituents (i.e. water ice VII and pure solid methane).
The contribution of the potential of interaction (in comparison to packing efficiency considerations) is found to be small and restricted to the lower pressure regime of the stability field.

Integrating Eq.(\ref{clausclap}) and following the dissociation curve for filled-ice, with respect to ice VII, a point of intersection with the melting curve for water ice VII is reached. Such an intersection is commonly referred to as a quadruple point. Up to the quadruple point the enthalpy difference between the $\alpha$ and $\beta$ phases is relatively small, as both phases are solid. Continuing the integration beyond the quadruple point the $\alpha$ phase will now represent fluid water. The enthalpy difference between the $\alpha$ and $\beta$ phases therefore increases, and becomes dominated by the enthalpy of fusion of filled ice. Unfortunately, the enthalpy of fusion of methane filled-ice Ih is experimentally undetermined and we estimate its value using the enthalpy of fusion for pure solid water at high pressure. 

\begin{figure}[ht]
\centering
\includegraphics[scale=0.6]{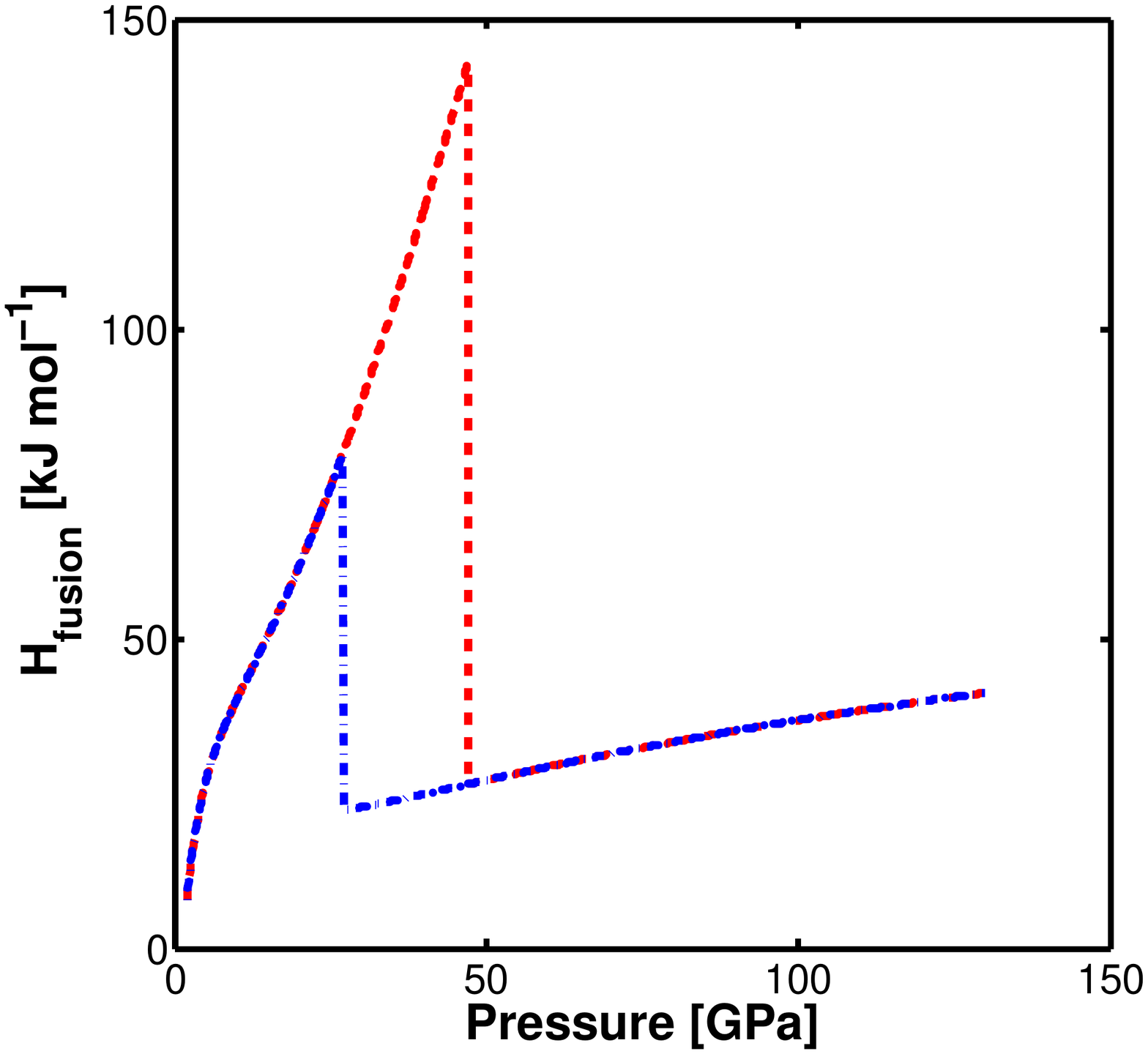}
\caption{\footnotesize{Dashed curve (red in the on-line version) is the enthalpy of fusion as a function of pressure for pure water \citep{Goncharov2009}. The branching in the melting curve, due to the super-ionic phase, is here at $47$~GPa \citep{Goncharov2005}. Dashed-dotted curve (blue in the on-line version) is a hypothetical enthalpy of fusion, where, the branching in the melting curve occurs at $27$~GPa.}}
\label{fig:Hfusion}
\end{figure}

\cite{Goncharov2009} deduced experimentally the melting curves for water ice VII and for super-ionic water. By further determining experimentally the volume difference between solid and melt, along the melting curve, they were able to derive the enthalpy of fusion for pure water. Their reported error is approximately $50\%$. As explained in \cite{Goncharov2009} the enthalpy of fusion increases with pressure since the increase in pressure means the melting transition is between a molecular solid and a fluid whose molecules become ever more dissociated and ionized. At a pressure of about $47$~GPa a branching occurs in the melting curve for pure water, due to the introduction of the super-ionic phase of water \citep{Goncharov2005}. The result is a reduction in the melting curve gradient and a sharp decrease of the enthalpy change upon melting \citep{Goncharov2009}. This behaviour is clearly seen in fig.\ref{fig:Hfusion}, where the dashed curve (red) reproduces the results from \cite{Goncharov2009} and the dashed-dotted curve (blue) is a hypothetical enthalpy of fusion, for which we have arbitrarily reduced the point of branching in the melting curve to $27$~GPa. 
The reason for the introduction of this hypothetical behaviour is that the enthalpy of fusion we adopt is from experiments on a homogeneous water system and therefore it is only an approximation for our methane filled-ice system. Molecular dynamic simulations by \cite{Iitaka2003} demonstrate that filled-ice exhibits behaviours similar to those appearing in homogenous water systems, but at somewhat lower pressures. A branching of the melting curve is therefore probable in our system of methane mixed with water as well, though, the branching may happen at a lower pressure then the $47$~GPa of the pure water system.
Wishing to test to what extent do our results depend on the point of branching we solve for the hypothetical enthalpy of fusion as well.

\begin{figure}[ht]
\centering
\includegraphics[scale=0.6]{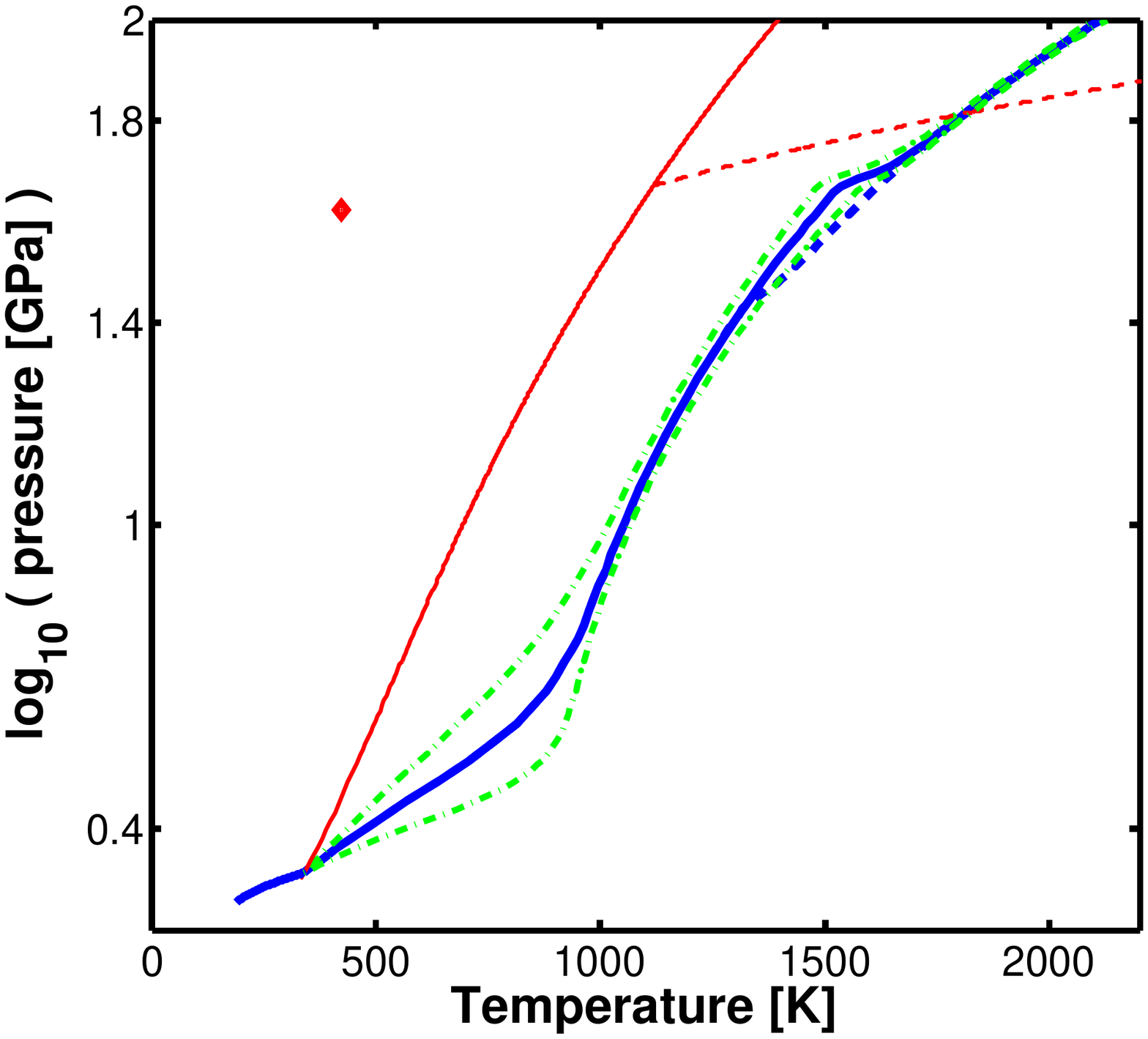}
\caption{\footnotesize{The solid curve (blue) is the phase diagram for filled-ice Ih assuming for it a thermal expansivity twice as much as that of water ice VII and a branching in the melting curve at $47$~GPa. The thick dashed curve (blue) is the variation in the phase diagram if the branching in the melting curve is shifted to $27$~GPa. The dashed-dotted thick curves (green) confining the thick (blue) curve from both sides represent the change in the phase diagram if the enthalpy of fusion is varied by $\pm 50\%$ globally. }}
\label{fig:depHfusion}
\end{figure}

In fig.\ref{fig:depHfusion} we solve Eq.(\ref{clausclap}) where we assume for filled-ice a thermal expansivity twice that for water ice VII and solve for each of the enthalpy of fusion scenarios depicted in fig.\ref{fig:Hfusion}. In the same figure we also show the solution of Eq.(\ref{clausclap}) with the experimental enthalpy of fusion from \cite{Goncharov2009} but vary it globally by $50\%$, which is its experimental error. From the figure we see that most of the variation in the filled-ice Ih dissociation curve due to the changes examined in the enthalpy of fusion occur below $10$~GPa.          

Combined with thermodynamic profiles for the interior of water exo-Earths, the thermodynamic stability field can help us decide whether filled ice structures can form at the core-mantle boundary and help convect methane and other hydrocarbons towards the surface.

\section{APPLICATION TO SUPER-EARTHS}
 
For the interior structure of water planets with masses up to 10 M$_{\oplus}$, \cite{fu10} consider a silicate-metal core, surrounded by a water-ice mantle. They find solid-state convection to prevail between the core-ice boundary and the conductive crust. Examining different surface temperatures and heat fluxes the authors find it possible for a liquid water ocean to exist beneath an ice-Ih solid surface layer. Different compounds expelled from the silicate-metal core may be incorporated in the ice matrix and convect outward. For the 2M$_{\oplus}$ planet \cite{fu10} find the silicate core-ice boundary pressure to range from approximately $50$ to $90$\,GPa, for water mass fractions between $25\%$ and $50\%$ respectively. The expected temperatures are between $700$ and $1000$\,K.

\begin{figure}[ht]
\centering
\includegraphics[scale=0.6]{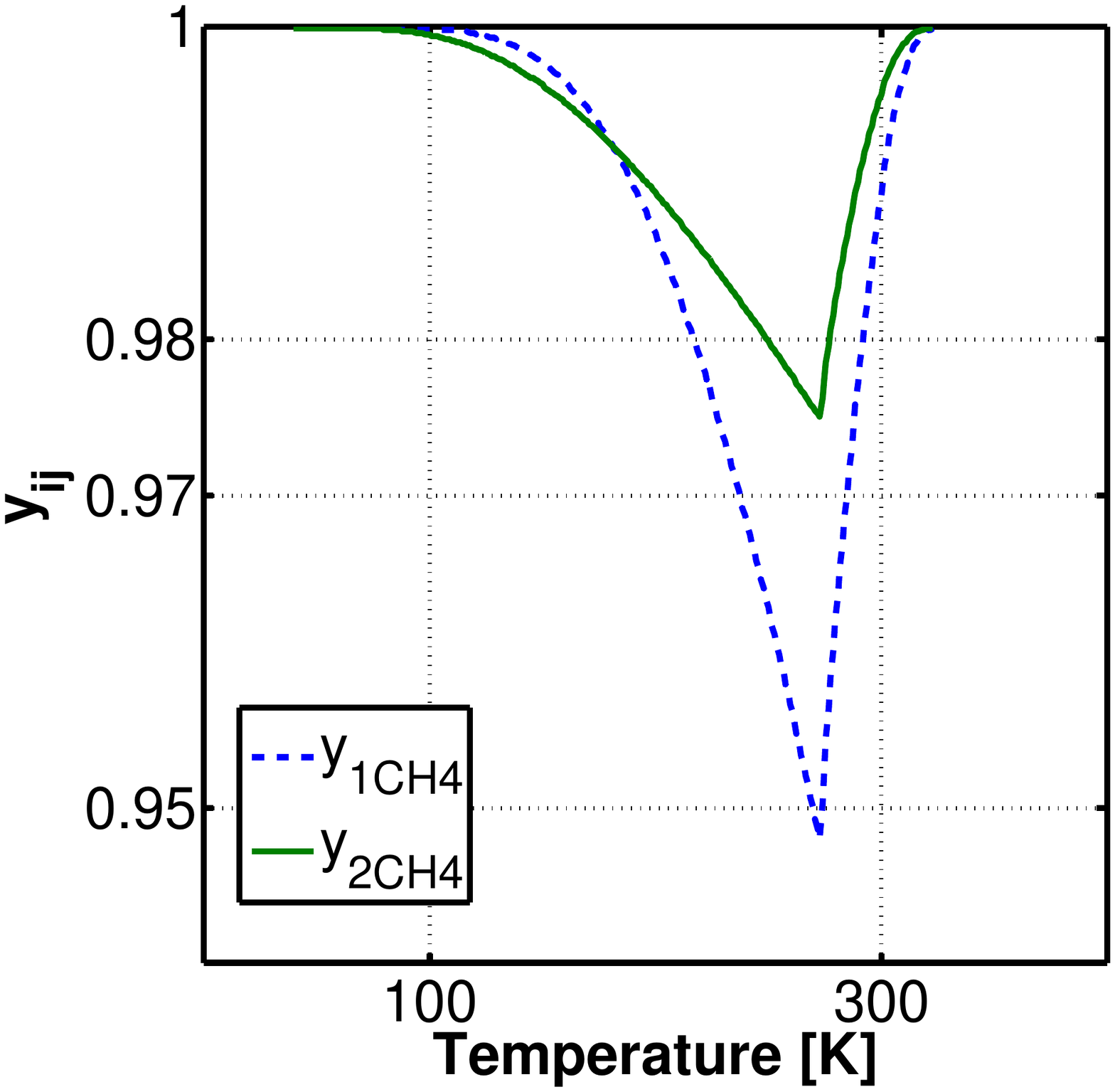}
\caption{\footnotesize{The probability of a methane molecule to occupy the large ($y_{2CH_4}$) and the small ($y_{1CH_4}$) cage of the structure I clathrate.}}
\label{fig:occupancy}
\end{figure}

Our extension of the theory of \cite{waalplat} suggests that the filled ice structure may be stable at such pressures and temperatures.  In such a case any CH$_4$ released from the core could be trapped in the lower part of the ice mantle. Experimentally \cite{loveday01} inferred a $2:1$ water-methane ratio for the filled ice phase.  As the mantle convects, the CH$_4$ would be carried to lower pressures where the filled ice will undergo a transition to a clathrate hydrate. Using the statistical model given by \cite{waalplat} we can compute the probability, $y_{Ki}$, that a structure I clathrate hydrate cage will be occupied.  This is shown in fig.\,\ref{fig:occupancy}.  It is clear that the occupancy is very close to full occupancy (probability of unity) for which there are $5.75$ water molecules per methane molecule. For a structure H methane hydrate the water-methane ratio ranges from 4.25:1 to 3.40:1, depending on the number of methane molecules occupying its large cage \citep{Koh2002}. 
Thus as the pressure decreases in an upwelling, excess CH$_4$ will be forced out of the water network. This may lead to the formation of a local methane reservoir.
Such reservoirs may naturally occur on the transition from filled ice to structure H clathrate hydrate at about $2$~GPa, and at the transition from structure H clathrate hydrate to structure I clathrate hydrate at about $1$~GPa. 

In A. Levi et al.(2013, in preparation) we show that the introduction of filled ice as a major constituent in an icy mantle has a large effect on the mantle thermodynamic profile. The probable higher thermal expansivity of filled ice compared with that for water ice VII results in a more moderate adiabatic thermal profile. The higher temperatures in the icy mantle, compared with those for a pure water mantle, creates a physical route through which super-Earths (objects less massive than Uranus and Neptune) may develop lower mantles in the super-ionic and reticulating phases. Phases so far related with the interior of bodies whose mass is equivalent to that of the icy giants of our solar system.
 
The unique characteristics of methane clathrate hydrate, namely its low thermal conductivity and the topology of its melting curve, yield water planets with thin crusts ($<$1~km) and a tendency to form a layer closely confined to the local melting condition beneath it. In that respect the geology and surface-atmosphere coupling in water super-Earths are quite different then simply assuming a water planet is a scaled up version of an icy satellite, such as Titan. Icy satellites tend to form thick crusts, in the order of $100$~km, and stabilize in a stagnant lid regime. This also means that chemical and geocycles in super-Earths are not simply resolved using analogies on icy moons.

The formation of a layer, in which conditions are close to melting, underneath the crust acts as a low viscosity layer that may decouple the convection cell and the crust, and represent a channel for relatively fast horizontal flow from up-welling to down-welling \citep{Crowley2012}. Such a fast horizontal flow may preferentially drag methane reservoirs to areas of down-welling where it may again become incorporated into the water matrix, preventing it from reaching the surface and the atmosphere.         
The time scale for transport of methane to the atmosphere is therefore broken down to several stages: the filled ice mantle overturn, the rate of transport in the low viscosity layer underlying the crust and the rate of collapse of the thin crust followed by exposure of fresh material. 

Other substances trapped in filled ice structures may be dragged preferentially along with the convecting ice. If they reach the top of the convection cell and meet with the bottom of a subsurface ocean, some compounds may favour clathrate-hydrate formation.  This will deplete those substances relative to others that are more easily transported in the liquid water and released to the atmosphere via cryovolcanism and surface ice tectonics. The volatile composition and abundance may also effect the ocean freezing temperature.  This will change the thickness and temperature of the solid crust. These issues are developed in more detail in A. Levi et al.(2013,  in preparation).

\begin{figure}[ht]
\centering
\includegraphics[scale=0.6]{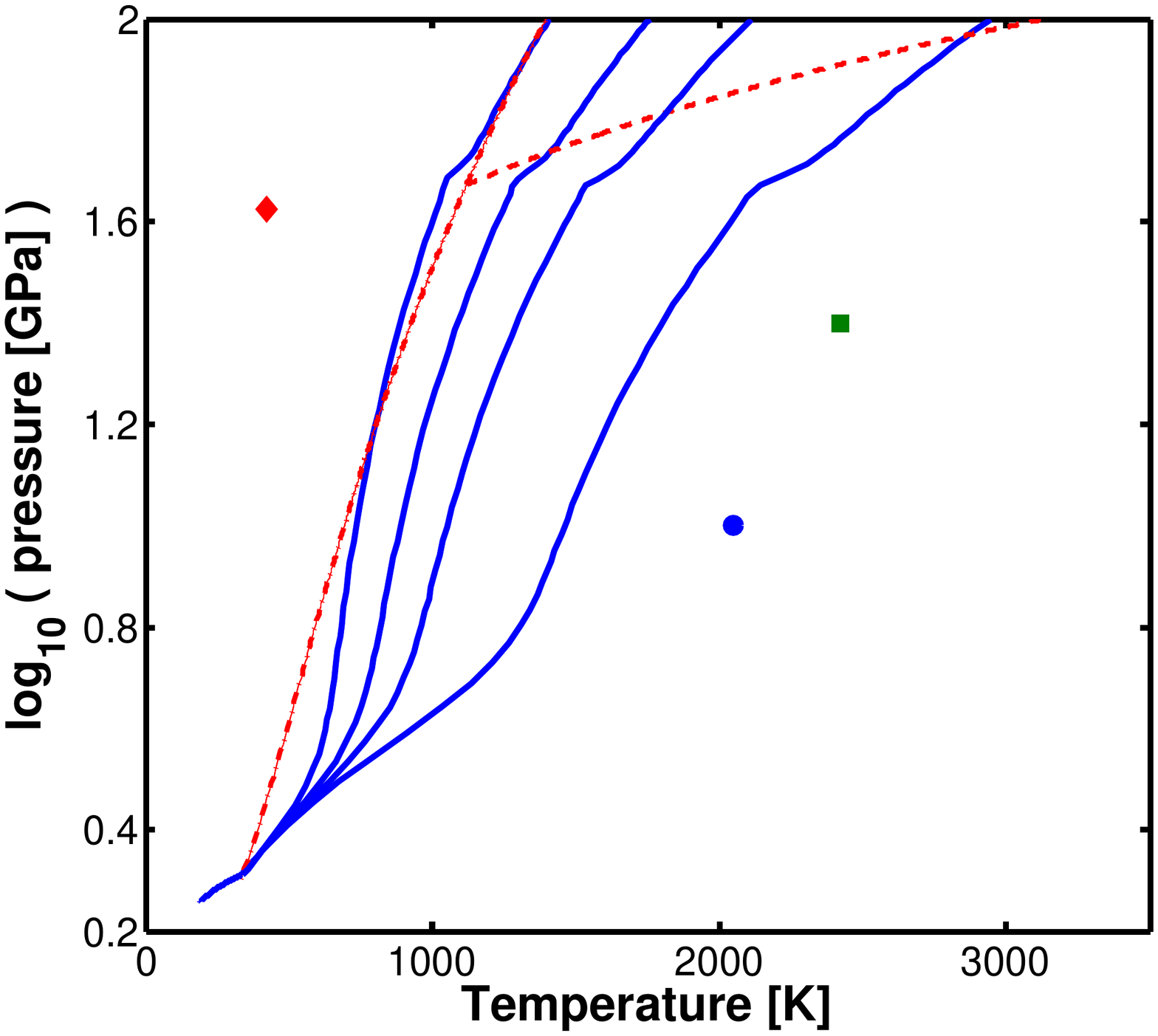}
\caption{\footnotesize{The phase diagram of methane filled ice-Ih water ice crystal. The diamond is a known point of stability \citep{hirai2003}. The four solid curves (blue in the on-line version) are the stability boundaries for the filled water ice, assuming for it a thermal expansivity equal to that for water ice VII and two, three and five times larger than that for water ice VII. A Lower thermal expansivity for filled water ice increases its stability, i.e. shifts the curve to the right to higher temperatures. The dash-dot curve is the melting curve for molecular water ice \citep{Lin2004} and the dashed curve is the melting curve for the super-ionic phase of water \citep{Goncharov2009}. The square and circle data points represent the conditions at the base of the H/He atmosphere surrounding Neptune and Uranus, respectively \citep{Redmer2011}. }}
\label{fig:phasediagram}
\end{figure}

In this work we have estimated the thermodynamic stability field for methane filled ice. From this field of stability it is clear that the blanketing effect of a thick H/He atmosphere, resulting in high internal temperatures, will most likely destabilize any filled ice (see fig.\ref{fig:phasediagram} for the P-T conditions at the base of the H/He atmosphere surrounding Uranus and Neptune). Water super-Earths lacking a substantial H/He atmosphere are more favourable for the formation of filled ice, which in turn will have a large influence on them. Additional experimental and theoretical work is needed to further constrain the properties of methane filled ice in order to better understand this influence. 

We wish to thank the anonymous referee for his constructive comments.       

\bibliographystyle{apj}
\bibliography{amitmemo} 

\end{document}